\documentclass[useAMS,usenatbib]{mn2e}
\usepackage{graphicx}
\usepackage{bm}
\usepackage{amsmath}

\title[A new method for computing self-gravity in an isolated system]{A new method for computing self-gravity in an isolated system}
\author[James H.H. Chan and Tzihong Chiueh]{James H.H. Chan$^{1}$ and Tzihong Chiueh$^{1}$$^{,}$$^{2}$$^{,}$$^{3}$\thanks{E-mail:
chiuehth@phys.ntu.edu.tw}\\
$^{1}$Department of Physics, National Taiwan University, 10617, Taipei, Taiwan\\
$^{2}$Institute of Astrophysics, National Taiwan University. 10617, Taipei, Taiwan\\
$^{3}$Center for Theoretical Sciences, National Taiwan University, 10617, Taipei, Taiwan}
\begin{document}

\date{}

\pagerange{\pageref{firstpage}--\pageref{lastpage}} \pubyear{2013}

\maketitle

\label{firstpage}

\begin{abstract}
\label{abstract}
A new approximation method for inverting the Poisson's equation is presented for a continuously distributed and finite-sized source in an unbound domain. The advantage of this image multipole method arises from its ability to place the computational error close to the computational domain boundary, making the source region almost error free. It is contrasted to the modified Green's function method that has small but finite errors in the source region. Moreover, this approximation method also has a systematic way to greatly reduce the errors at the expense of somewhat greater computational efforts. Numerical examples of three-dimensional and two-dimensional cases are given to illustrate the advantage of the new method.
\end{abstract}

\begin{keywords}
gravitation – gravitationallensing:strong – methods:numerical
\end{keywords}

\section{Introduction}
\label{sec:introduction}
Inversion of Poisson's equation with a finite-size source in an unbound system is commonly encountered in Physics, Astronomy, Quantum Chemistry \citep{bSzab} and Fluid Mechanics \citep{bGunz}.  Examples include the electrostatic problems, magnetostatic problems, gravity problems, etc.  With an arbitrary source distribution in an infinite space, one would like to know what the potential appears in space and how the resulting force may act back to the source.  For incompressible flows in fluid mechanics, pressure also satisfies the Poisson's equation that needs to be solved in order to evolve the flow.   Here the source of the Poisson's equation is related to the flow vorticity, which is often localized within the active region of interest in an unbound domain.  Another type of problems involves the Monte Carlo search, such as gravitational lens problems. Here one needs to search for various lens mass distributions to obtain images consistent with observations.  

For dynamical or Monte Carlo problems mentioned above, the Poisson's equation must be solved at every time step to properly follow the evolution or solved for every trial to arrive at the global minimum. A fast and accurate Poisson's solver is therefore desired. In the past couple of decades, a fast and accurate Poisson's solver, the modified Green's function method, has already been available for a finite-size source in an unbound space \citep{bEast, bFell, bBudi}. The modified Green's function method properly takes into account the image charges outside the computational boundary so that the solution satisfies the periodic boundary condition and can be correctly evaluated within the original domain. Moreover, the Fast Fourier Transform (FFT) can be adopted in this method for speedy computation.  However, the cost of this method is a large computational domain required to include the contributions from image charges. The larger volume is $2^D$ times the original volume, where $D$ is the spatial dimension. Despite that one may reduce the computation by taking advantage of some symmetries in the Green's function, the $2^D$ scaling is unavoidable. As a result, this method begins to lose its computational speed edge for high-dimensional problems.  

However, the modified Green's function method actually contains non-negligible errors in regions where the sources are located, especially the source has a large gradient.  The error is originated from the assignment of an ambiguous value for the modified Green's function at the very grid where the point source is located. Different values have been proposed for different situations \citep{bFell}, and there has been no universally agreed good choice. 

In this paper, we provide the motivation of the proposed method in Section \ref{sec:motivations}. We then give the step-by-step recipe of the new method in Section \ref{sec:detailed correction procedures}. A three-dimensional example and a two-dimensional example are provided to illustrate the performance of the new method in Section \ref{sec:numerical examples}. Section \ref{sec:discussion} discusses a possible extension and concludes this work.

\section{Motivations of The New Method}
\label{sec:motivations}

The numerical errors of the modified Green’s function method are mostly located around where high numerical accuracy is desired. For the same amount of errors, if they were to be located in regions where the sources are absent, the solution accuracy would have been greatly improved. This is the primary motivation behind our new method. The errors produced by the new method will be moved close to the boundary where the source is nearly absent. Our secondary motivation is the computational speed.  We seek a new method that computes the force within the original volume with FFT, and in a three-dimensional calculation it can save the computation by a factor of at least two as compared with the modified Green's function method.  However, this advantage may not be valid in low dimensional problems, such as the two-dimensional gravitational lens problem, where the saving in computation can be limited.

We now further substantiate our primary motivation. When we compare the forces computed from a isolated source and those computed by FFT from the same source, two types of error sources are found, a long-range image monopole error extending over the whole domain and short-range image multipole errors near the domain boundary.   Take the gravitational problem as an example. The monopole error arises from the attraction of all image masses, which correctly produces a null force at the mass center, but the force error grows away from the mass center. To remove error produced by the monopole images, we can treat the distributed mass as a mass point located at the mass center, and it is conceptually straightforward to subtract off the erroneous forces given by all image mass points. There is no dipole contribution for gravity and so the next order error is from the quadruple moments of image masses, which produce far-field short range forces, proportional to $r^{-4}$ from the image positions, and the largest errors are near the domain boundary.    The far-field force errors produced by the image quadrupole moments can similarly be also subtracted off conceptually, if we know how to sum the image contributions.   The correction procedure can continue to any arbitrary multipole order systematically.  After the $N$-th order corrections the remaining error force will be ever shorter range, $\propto 1/r^{N+3}$, which decays fairly rapidly across the boundary into the domain.

Once the principle for error subtraction is clear, the remaining question is how to sum up the far-field multipole forces accurately from infinitely many images.  We may take each conventional multipole moment expansion of the original density distribution as the source, use FFT to invert the Poisson's equation, and compute the force that includes all image contributions, which we call the multipole FFT force.  The desired error sum is obtained by subtracting the FFT force from the exact multipole far-field force of the actual density distribution. However, this procedure will yield large numerical errors for the FFT force near the mass center because the multipole forces are singular ($\propto r^{-(l+2)}$) at the mass center. 

To circumvent this problem, a different approach is adopted.   We deliberately design a well-behaved template multipole moment density distribution, for which the multipole force is also well-behaved everywhere and has an exact analytical expression. This template multipole density distribution has the same far-field multipole moment as the original density distribution, and is then inverted to obtain its FFT force.  The desired image sum is obtained by subtracting the FFT force from the exact multipole force, and this is the force error to be removed. Such a procedure avoids singularities at the mass center and is able to be carried out order by order in multipole expansion.  

\section{Detailed Correction Procedures}
\label{sec:detailed correction procedures}

        Below, the procedure for image multipole correction for three-dimensional case is described:

(a)	The multipole moment of the source, defined as $M_{lm}=\int \rho({\bf r})r^{l}Y_{lm}(\theta,\phi) d^3{\bf r}$, is first computed up to the order of correction desired.  

(b)	A template of multipole density $a_{lm} Y_{lm}(\theta,\phi) T_{l}(r)$ with analytical expressions of force ${\bf F}_{lm}^T(\bf{r})$ is given, and we let $a_{lm}$=$M_{lm}/\int Y_{lm}(\theta, \phi) T_{l} (r) d^3r$ so that the template has the same multipole moment as the original density.

(c)     The Poisson's equation is then inverted via FFT in the original domain with the template $a_{lm} Y_{lm}(\theta,\phi)T_{l}(r)$ as the source to compute the multipole FFT force, ${\bf F}_{lm}^{FFT}({\bf r})$.   The difference between the FFT force and the known exact analytical force, $\Delta {\bf F}_{lm}({\bf r})={\bf F}_{lm}^{FFT}({\bf r})- {\bf F}_{lm}^T(\bf{r})$, then yields the correction force needed for this particular multipole moment. In fact, once every $a_{lm}$ is determined, all $a_{lm} Y_{lm}(\theta,\phi)T_{l}(r)$ can be summed together before the FFT inversion.  After the overall FFT force is obtained, it is then subtracted from the overall exact analytical force to determine the overall force correction.

(d)	The FFT force obtained from the original density distribution is finally subtracted from the result of (c).  The corrected forces will have high accuracy in the control region of the computational domain with under-corrected errors confined near the domain boundary.

For a two-dimensional problem, similar procedures can be straightforwardly followed. As a technical note, each multipole component of the template multipole density and force has a symmetry. When any of the seven mappings $(x\to-x)$, $(y\to -y)$, $(z\to -z)$, $(x,y\to -x,-y)$, $(x,z\to -x,-z)$, $(y,z\to -y, -z)$ and $(x,y,z\to -x,-y,-z)$ is performed, they will assume the same values but with different signs given by the symmetry.  It then follows that the template multipole densities and forces need to be computed only in $1/8$ of the domain.  When the computer memory space is allowed, each template multipole density and force should be computed only once and stored in the memory, as they will always be the same regardless of the changing source mass distribution.

\section{Numerical Examples}
\label{sec:numerical examples}
In Appendix, we list the template density-potential pairs used in these examples.  

(a)	Three Dimensional Case ---  Gaussian Mass Spheres:

We provide a 3D example with 6 Gaussian spheres of the same central densities but various sizes to demonstrate the accuracy of the present method compared with the modified Green's function method.  Each Gaussian sphere has an exact known force and the composite forces from these Gaussian spheres are also known exactly by superposition.  The sizes of the Gaussian spheres vary from $\sigma=3$ grids to $\sigma=30$ grids in a domain of $512^3$ grids. For convenience of displaying the errors produced by the modified Green's function method, which are concentrated at the sources, we deliberately place the centers of six sources on two orthogonal planes; on each plane, 3 sources are placed randomly but confined within the inner half of the box. One plane contains 3 narrow spheres and the other 3 wide spheres. 

\begin{figure}
\includegraphics[scale=0.35, angle=270]{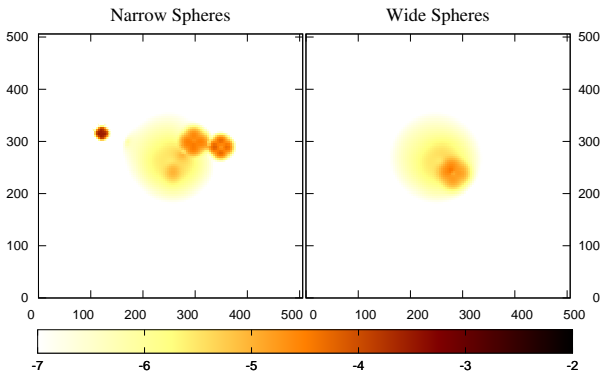}
\caption{Two slices cut through the 3D sources to show the ratio of the force error relative to the original force computed by the modified Green's function with the color bar in the base 10 logarithm scale. The 3 gaussian spheres vary from $\sigma = 3\sim8$ pixels in the left one, and the other 3 vary from $\sigma = 10\sim30$ pixels in the right one. The coordinates are in pixels. Note that the error is larger at the locations where the sources are placed.}
\label{fig:err_green}
\end{figure}

\begin{figure}
\includegraphics[scale=0.35, angle=270]{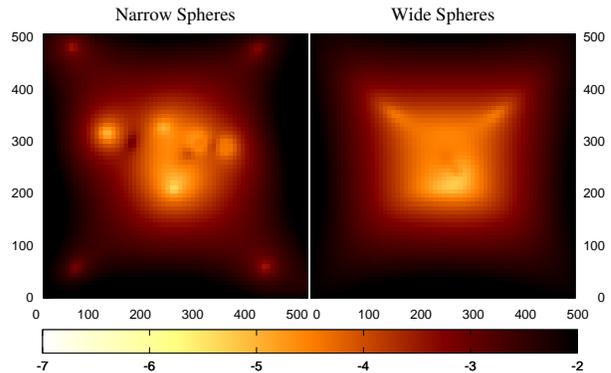}
\caption{Same as Fig. (\ref{fig:err_green}), except for the use of the image multipole method. Noted that the error is larger around the peripheral region.}
\label{fig:err_multipole}
\end{figure}

The metric of goodness for any given method is the ratio of the residue error force strength to the original force strength at the same location. To avoid the force error arising from numerical differentiation on the potential produced by the modified Green's function method, the three force components are also directly computed by analytically differentiating the modified Green's function. We then examine the residue error configurations for both the modified Green's function method and the image multipole method. The multipole image method contains multipoles up to $l=4$.  Plotted in Fig. (\ref{fig:err_green}) is the percentage error of force for the modified Green's function method, and in Fig. (\ref{fig:err_multipole}) that for the image multipole method. These errors are displayed on the two slices where the source centers lie.   It is clear for the modified Green's function method that the force errors are confined within the sources, with smaller spheres having larger force errors, and away from the sources the force error rapidly approaches zero. The error is of small scale and can lead to internal distortion of the mass objects, more so for smaller ones. On the other hand, the errors peak at the boundary for the image multipole method, with the error strongly depending on the distance from the boundary and being insensitive to the detailed spatial distribution of the sources. These errors are smooth and of large scale, and can lead to erroneous acceleration of the outlier objects, more so for those closer to the domain boundary.

\begin{figure}
\includegraphics[scale=0.35, angle=270]{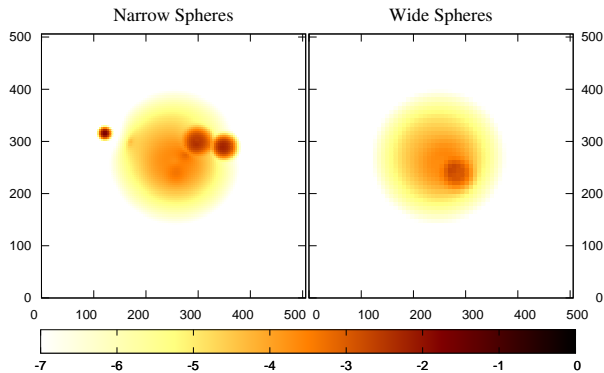}
\caption{Same as Fig. (\ref{fig:err_green}), except for the use of the Green’s function that has no tuning of radial derivatives near the central singularity.  Note that the error is almost two orders of magnitude greater compared to Fig. (\ref{fig:err_green}).}
\label{fig:err_green_s}
\end{figure}

We however note a caveat for the modified Green's function method.  The accuracy of this method depends on the value of the radial derivative (force) immediately next to the central singularity of the modified Green's function. This feature is similar to a $r$-dependent softening length $\epsilon(r)$ in the force, $1/(r^2+\epsilon(r)^2)$, where the profile of the softening length depends on the size of the mass object.  We find that a particular size of mass object is optimized by one particular value of force around the central singularity, and this value is not optimal for objects of other sizes.  Normally the optimal force is larger for a more compact object.  Even after optimization, the most compact object still contains the biggest error.  In the present case, we take the Green's function force one pixel away from the singularity to be $1.48$, to be compared with the force value $1$ given by the unmodified force law $1/r^2$ at the same location $r=1$.  This larger value of Green's function force, representing force hardening, minimizes the force error in the most compact object.  For a comparison with Fig. (\ref{fig:err_green}), shown in Fig. (\ref{fig:err_green_s}) is the much greater percentage error by having all pixels to follow the unmodified force law $1/r^2$.   It is surprising to find that one actually needs the force hardening to capture the correct gravity of a continuous source instead of the force softening used to mimic the softened gravity of collisionless particles. Fig. (\ref{fig:err_green_s}) also shows that the errors are isotropic compared with the octuple errors in Fig. (\ref{fig:err_green}) using the optimal parameter.  This result reflects the fact that minimization of force errors is often incompatible with isotropy of force errors. For some applications it is worth sacrificing some force accuracies for a more isotropic force.
  
Despite that this paper aims to address the fluid-based self-gravity, which has subtle differences from the particle-based self-gravity, we briefly address the particle-based gravity to shed light on its similarities. In astrophysics, particles are normally collisionless, and to avoid two-body relaxation a soften length is implemented, thereby smoothing the singular gravity of a particle. The degree of softening depends on particle density, with more softening in a system of less particles density \citep{bDehnen}. In fact, by carefully shaping the particle softening, one can achieve the force resolution at the grid scale and below. The continuity of force resolution extending below the grid scale permits the addition of sub-grid particle-particle interactions, which is the basis for the Particle-Particle-Particle, Mesh method aiming to solve high density gradient problems. The fluid-based self-gravity is the limiting case for infinitely many particles per cell, therefore requiring minimal force softening; in fact, the optimal case turns out to require force hardening.  Also, compared with the modified Green's method the image multipole method performs well for small-scale structures, as illustrated in Figs. (\ref{fig:err_green}) and (\ref{fig:err_multipole}), thus capable of handling high source gradient problems. 

The absence of an optimal modified Green's function for general mass distributions makes it impossible to evaluate the force errors accurately. Nevertheless the force error generally scales as $(\Delta x/d)^3$ for a given choice of Green's function central derivative, where $d$ is the size of the mass object and $\Delta x$ the grid size, as described below.  On the other hand, the force error for the image multipole method depends on the truncation degree, $l+1$, of the multipole expansion for sources located at least half box size $L/2$ away, and hence the largest force error at the boundary for the image multipole method scales as $(L/2)^{l+3}$.  To test how each method can be improved, we increase the grid resolution by a factor of two for the Green's function method and enlarge the domain by a factor of two to place the boundary twice further away for the image multipole method. Both increase the pixel number by a factor of 8.  As expected, the latter improves by two orders of magnitude since the residue force error arises from the contribution of $l=5$ and beyond.  But the former improves only by a factor about 8.    This comparison illustrates that the image multipole method has a systematic way for drastic improvements whereas the modified Green's function method has only limited improvements. 

\begin{figure}
\includegraphics[scale=0.35, angle=270]{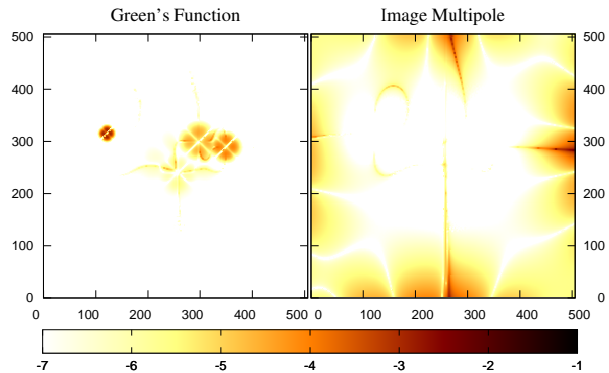}
\caption{The geometric errors $|\partial_x\delta F_y-\partial_y\delta F_x|$ of the three small spheres computed by both methods normalized to the corresponding component of the shear tensor $2|\partial_x\partial_y\phi|$, where $\phi$ is the potential. The color bar is also in the base 10 logarithm. Again, the error distributions are similar to force errors in Figs. (\ref{fig:err_green}) and (\ref{fig:err_multipole}).}
\label{fig:err_curl}
\end{figure}

To better illustrate the problem of the modified Green's function method, we now consider the geometric error $\nabla\times {\bf F}$, which must be exactly zero for the gravitational force.   As a result of force errors, the numerical gravitational force can contain a pseudo-vector. Given the force errors of Figs. (\ref{fig:err_green}) and (\ref{fig:err_multipole}), we now show $\nabla\times \delta{\bf F}$ yielded by both methods in Fig. (\ref{fig:err_curl}). For convenience, we show only one component of $\nabla\times \delta{\bf F}$ perpendicular the plane of slice in Fig. (\ref{fig:err_curl}). To gain an idea of the magnitude of the error, we normalize $|\partial_x\delta F_y-\partial_y\delta F_x|$ by $|\partial_x F_y+\partial_y F_x|=2|\partial_x\partial_y\phi|$, a component of the zeroth-order shear tensor, where $\phi$ is the potential.  It is clear that the modified Green's function method creates substantially large geometric errors interior to the domain, especially concentrated at locations where the mass objects are. It can produce non-negligible erroneous circulation flows within mass objects.

Finally, let us turn to the computational speed for the gravitational force calculation, where the inversion of Poisson's equation may be a bottleneck for 3D dynamical simulations.   We find the force calculation of the modified Green's function method takes about twice more time than that of the image multipole method.  This ratio can be accounted for by the fact that the modified Green's function method computes FFT and inverse FFT in a volume 8 times the original volume, but the image multipole method computes FFT and inverse FFT two times in the original volume, one for the actual mass distribution and the other for the template mass distribution.  This gains a factor of 4 in the FFT computations for the image multipole method. The remaining extra tasks to compute the moments and the template forces only take an extra small fraction of FFT computing time to come up with a speedup factor 3.6. On the other hand, the Green's function method can take advantage of the symmetry of the spectral space to reduce the computation by roughly a half. Thus, the image multipole method is a factor 1.8 times faster than the modified Green's function method. 

(b)	Two dimensional case --- Lens within Lens:

A practical two-dimensional example for a strong gravitational lens problem \citep{bBlan, bKorm, bKeeta} is considered here. We consider two cored isothermal lenses \citep{bKorm, bKeeta} at the same redshift.  The main lens is elliptical located at the center and the small lens, as a subhalo, is spherical located near the critical line \citep{bBlan} of the main lens; thus the critical line of the main lens gets seriously distorted.   It is quite common in the lens observations that one of the multipole images is located near a small lens \citep{bSuyu, bVege, bRich}, and is therefore perfect for illustrating the strength of the image multipole method.   

\begin{figure}
\includegraphics[scale=0.38, angle=270]{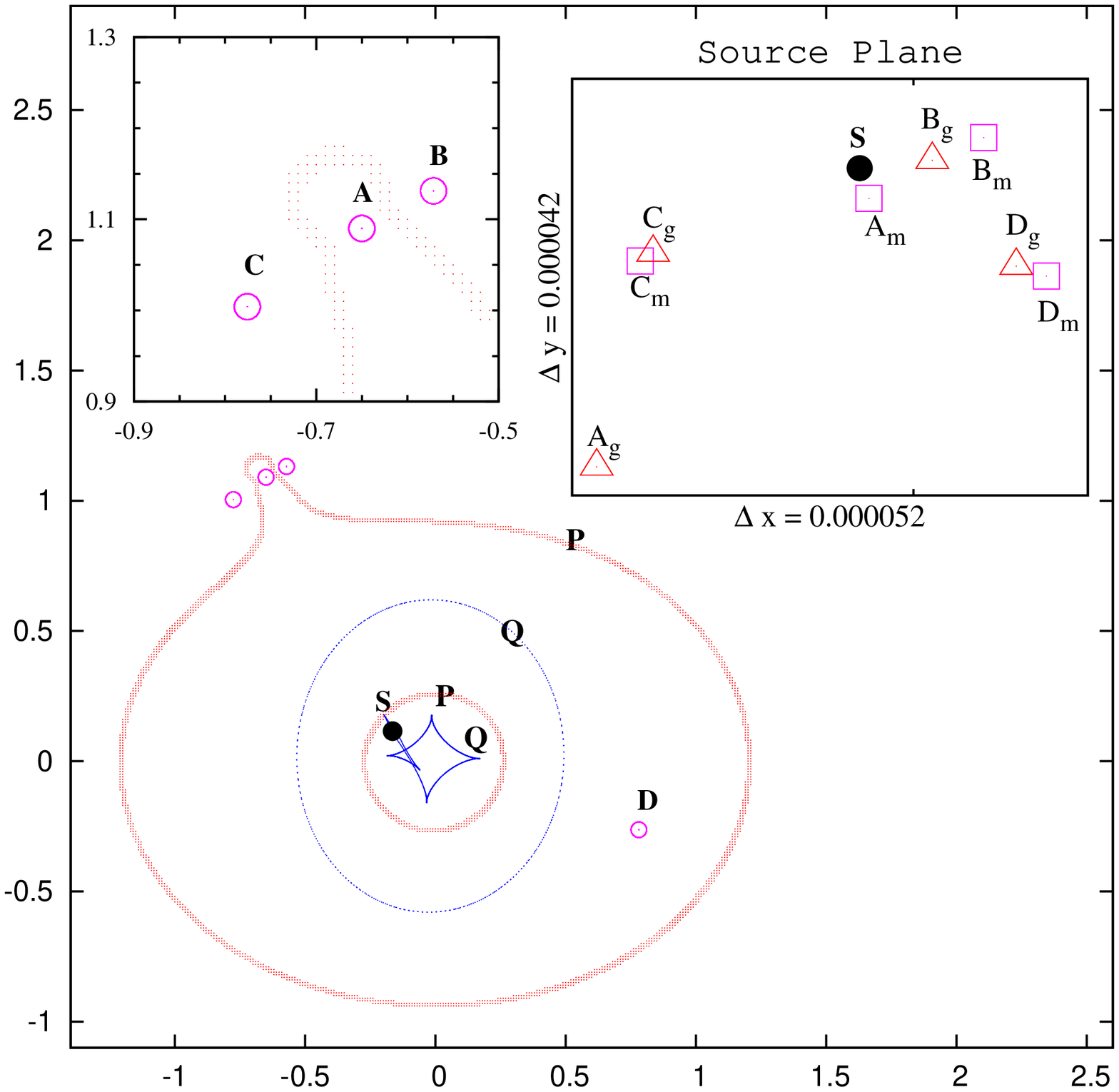}
\caption{Quasar quad images caused by gravitational strong lensing by one elliptical lens at the center and a smaller spherical lens near the upper left of the main lens critical line(P). The source labeled as "S" is located near the caustics(Q) and produces four images(empty circle). The coordinate is in arcsecond. The top right inset shows the zoom-in ($0.000052\arcsec\times 0.000042\arcsec$) view of the four source positions produced by the modified Green's function(triangle and subscript g) and image multipole(square and subscript m) methods.} 
\label{fig:lens_ex}
\end{figure}

Shown in Fig. (\ref{fig:lens_ex}) is the critical line of radius about 1$\arcsec$, corresponding to a $10^{12}M_{\odot}$ lens galaxy at $z=0.5$ and a source at $z=2$.   The critical lines produced by the two numerical methods and by the analytical solution are indistinguishable from each other at the resolution of Fig. (\ref{fig:lens_ex}).  We let the source (black dot) be a quasar, for which the optical emission region is on the order of one micro-arcsecond, i.e., practically a point source.  Moreover, the source is located near the corner of the caustics and produces a quad image, where one (A) of the three near-by images (B, A and C) is co-spatial with the small lens, as depicted in the upper-left inset. Note that the forces of elliptical and spherical core isothermal lenses have analytical expressions, and hence we know the exact loci of caustics and critical curves.

\begin{figure} 
\includegraphics[scale=0.5, angle=270]{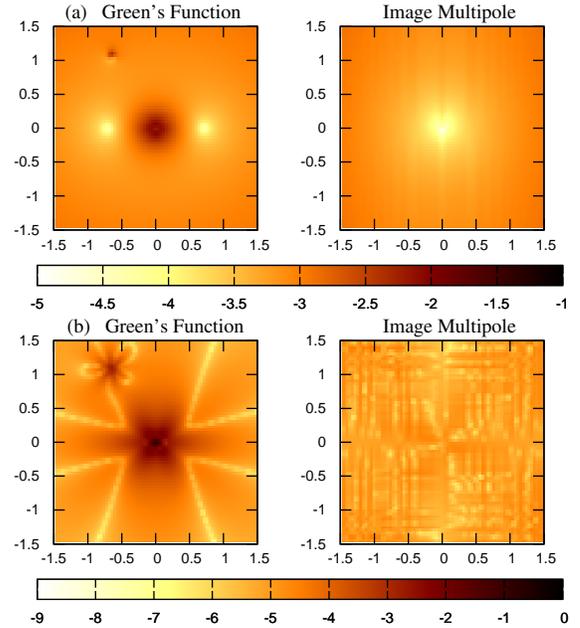}
\caption{The errors of deflection angles (a) and magnifications (b) computed by the modified Green’s function (left) and image multipole (right) methods.  The coordinate is in arcsecond, and the color bar is in the base 10 logarithm.}
\label{fig:err_compare}
\end{figure}

In the lens mass modeling, one normally obtains four source positions calculated from the quad image positions for a given lens mass model.  Even for a fairly accurate mass model, it is nearly impossible to yield four spatially coinciding sources.  As a result, one evaluates the accuracy of the lens mass model by finding how close the four sources are from the best source position, and how well the magnifications of the four images agree with the observed fluxes.  The upper-right inset of Fig. (\ref{fig:lens_ex}) shows the zoom-in view of the four source positions produced by each method.  In this case, we have used $1024^2$ pixels in the image plane for analysis, with a field 5 times bigger in linear size than that of the Einstein ring of 1$\arcsec$ radius. This corresponds to 0.01$\arcsec$ per pixel.  In addition, image multipoles are corrected up to $m=6$.  Before addressing the error estimate $\chi^2$ of each method, we show in Figs. (\ref{fig:err_compare}a) and (\ref{fig:err_compare}b) the absolute values of errors in deflection angles($\propto\nabla\phi$) and in magnification $1/[(1-\kappa)^2-\gamma^2)]$. Here, $\kappa$ is the surface density normalized to the critical surface density and $\gamma$ is the shear $|\partial_i\partial_j-(\delta_{ij}/2)\nabla^2)\phi)|/2\pi G$ normalized to the critical surface density \citep{bSchn}, where $\delta_{ij}$ is the Kronecker delta function.  The performance differences in deflection angles and in magnifications are obvious. 

Whether such errors are acceptable or not is gauged by comparing these errors with observational uncertainties, namely the positional uncertainty (seeing blurring and telescope diffraction) and the flux uncertainty (photometry error), both of which are evaluated at the image plane.  However, the positional errors are analyzed at the source plane in practice. To account for this difference, we divide the image positional uncertainties by the squared root of magnification of every individual image to translate the image positional uncertainties to the source positional uncertainties, for the reason that a blurred image after de-lensed should recovered a sharp source with a smaller uncertainty.  Also, as the four quasar images have different brightness, therefore different signal-to-noise ratios for determination of image positions, we hence optimally weigh each image position by the square root of magnification again in the definition of $\chi_r^2$ (see below).    

Having such an adjustment for the error measure, we now consider the Hubble Space Telescope observations of the ultra-violet channel, for which the positional uncertainty $\sigma_r\sim 0.0004\arcsec$ and the photometry uncertainty $\sigma_m=0.02$ magnitude.  We let $\chi^2$=$\chi_r^2 +\chi_\mu^2$, the sum of contributions from positional errors and magnification errors, where 
\begin{equation}
\label{equ:chi2_src}
\chi_r^2= \sum_i{\mu_i^2({\bf r}_{src}^i-{\bf r}_{src})^2\over\sigma_r^2}
\end{equation}
and
\begin{equation}
\label{equ:chi2_flux}
\chi_\mu^2=\sum_i{((f_i/f_{src})-\mu_i)^2\over\sigma_m^2},
\end{equation}
with ${\bf r}_{src}^i$, and ${\bf r}_{src}$ being the calculated source position and the exact source position, and $f_i$ and $f_{src}$ the image flux and the expected source flux \citep{bKoch, bKeetb, bKopt} and $\mu_i$ is the exact magnification of each image.   Here the index $i$ runs from 1 to 4 for the quad images.  The example configuration given in Fig. (\ref{fig:lens_ex}), quite common in strong lens observations, yields that $\chi_r^2=1.08$ and $0.3$, and $\chi_\mu^2=972$ and $2.5$ for the Green's function method and the image multipole method, respectively.  Not surprisingly, the magnification errors set it apart for the performance of the two methods; the image multipole method produces far more accurate image magnifications.   Even for the positional errors, the image multipole method is still better by a factor 3.

\section{Discussion and conclusion}
\label{sec:discussion}
There are a couple of caveats for the image multipole method.   First, when the mass distribution is extended near the computational boundary, the domain of computation must be substantially enlarged for this method to work properly, thereby increasing the computational load.   This issue on the other hand creates lesser a problem for the Green's function method, as it works even when the mass extends up to very close to the boundary.   In such a case, the image multipole method will lose the edge. This difficulty cannot be easily alleviated by increasing the order of truncation.   The increase of multipole order from $l$ to $l+1$ requires $2l+3$ more volume integrations for evaluating the multipoles, with a gain in the error reduction by a factor only of order $d/L$, where $d$ and $L$ are the source size and the computation box size, respectively.   If $d$ is comparable to $L$, there is essentially no gain compared to the strategy of increasing the box size, for which the force error is reduced by $(L/L')^{(l+3)}$, with $L$ and $L'$ being the original and the enlarged computational box sizes.   

A related problem is when the source does not vanish at the computational boundary, such as an infinitely-extended Plummer's sphere.  The mass distribution of the Plummer's sphere in an isolated cube contains not only the monopole moment but also $l$ moments beyond $l=2$.  Hence the force is no longer radial.  The image multipole method cannot recover the ideal radial force of Plummer's sphere but the force of mass distribution containing $l=0$ and even $l$ moments beyond $l=2$.  This problem is generic and also exists for the modified Green's function method.  To avoid this problem, the source must be truly isolated and vanishes at the computational boundary.

Second, when the mass distribution contains no substantial low-order multipole moments other than the monopole, the image multipole method must proceed to sufficiently high orders to exercise its corrective power.   The situation occurs for very symmetrical mass configurations, for example, 6 identical mass clumps located at the center of the 6 faces of a cube, which, apart from the monopole, has multipole moments beginning at $l=4$, and the corrections for $l=2$ and $l=3$ have null effects.   This is probably the worst scenario for the image multipole method. 

In view of these problems for the image multipole method, it may be possible to adopt a hybrid method combining the strengths of both.   Near the domain boundary, one can use the modified Green's function method to compute the isolated gravity, and in regions where mass clumps are present, one can employ the image multipole method.   Matching the gravities computed by both methods at locations where errors of both are small can be a non-trivial problem.   But given the opposite trends clearly shown in this work, the hybrid method deserves serious attention when a highly accurate solution of Poisson's equation is desired.  

In sum, we present a new method to compute the force given by a finite-sized source of the Poisson's equation.  This new method places the numerical errors close to the boundary, leaving the source region almost error free.   The performance is compared with an existing method, the modified Green's function method.  With the image multipole correction up to $l=4$ for 3D and $m=6$ for 2D, we show that the image multipole method can create smaller errors at the source region, and moreover in 3D calculation its computation load can be a factor of $2$ lighter than the modified Green's function method.  Unlike the modified Green's function method, where systematic improvements of the approximation are quite limited, the accuracy of this new method can be increased drastically by enlarging the computational domain, making this method generally a better choice when high numerical accuracy is desired.

\section*{Acknowledgements}
We acknowledge the support from the National Science Council of Taiwan with the grant: NSC-100-2112-M-002-018. We thank Ui-Han Zhang for many useful discussions.

\appendix
\section{Template Density-Potential Pairs}
The choice of density-potential pairs is not unique, provided the density is regular and nearly zero close to the domain boundary. Here we show the template profile adopted in this paper as an example. For the 3D case, our density-potential pairs for multipoles are defined as
\begin{equation}
\label{equ:phi_defn}
\phi_{lm}(r,\theta,\varphi)=a_{lm}\Phi_{l}(r)Y_{lm}(\theta,\varphi)
\end{equation}
and
\begin{equation}
\label{equ:rho_defn}
\rho_{lm}(r,\theta,\varphi)=a_{lm}T_{l}(r)Y_{lm}(\theta,\varphi),
\end{equation}
where 
\begin{equation}
\label{equ:poisson_3d}
{1\over r^2}{d \over dr} \left( r^2 {d \Phi_{l}(r) \over dr} \right) - { l(l+1)\over r^2}\Phi_{l}(r)=4\pi G T_{l}(r).
\end{equation}
Specifically, we choose the radial dependence of the monopole as
\begin{equation}
\Phi_{0}(r)=-{1\over r}{\mathrm{Erf}}\left({r \over \sqrt{2} a}\right)
\label{equ:rho_3d_m}
\end{equation}
and
\begin{equation}
\label{equ:phi_3d_m}
T_{0}(r)={1\over{\left(a\sqrt{2\pi}\right)}^3} e^{-{r^2\over 2a^2}},
\end{equation}
and the radial dependence of the higher order multipoles as
\begin{equation}
\begin{split}
\label{equ:rho_3d_q}
\Phi_{l}(r) = { {(1-e^{-{r^2\over 2b^2}})}^{2l+1}\over r^{l+1}}
\end{split}
\end{equation}
and
\begin{equation}
\begin{split}
T_{l}(r) = & {\left(2l+1\right)e^{-{r^2\over 2b^2}}{\left(1-e^{-{r^2\over 2b^2}}\right)}^{2l-1}\over b^2 r^{l+1}} \times\\
& \left[{2lr^2 e^{-{r^2\over2b^2}}\over b^2} - \left(1-e^{-{r^2\over2b^2}}\right)\left(2l-1+{r^2\over2b^2}\right)\right].
\end{split}
\label{equ:phi_3d_q}
\end{equation}
In this work, we choose $a=10$ and $b=20$ for the 3D template profile. The unit of $a$ and $b$ is in pixel.
Similarly for this radial profile for the 2D case, we define
\begin{equation}
\label{equ:phi_2d_defn}
\phi_{m}(r,\varphi)=b_{m}\Phi_{m}(r)e^{im\varphi}
\end{equation}
and
\begin{equation}
\label{equ:rho_2d_defn}
\rho_{m}(r,\varphi)=b_{m}T_{m}(r)e^{im\varphi},
\end{equation}
where
\begin{equation}
\label{equ:poisson_2d}
{1\over r}{d \over dr} \left( r {d \Phi_{m}(r) \over dr} \right) - { m^2\over r^2}\Phi_{m}(r)= 2 \pi G T_{m}(r).
\end{equation}
The radial dependence of the monopole is chosen to be
\begin{equation} 
\label{eqn:phi_2d_m}
\Phi_{0}(r)= {1\over 4\pi} \ln\left(e^{-{r^2\over2a^2}}+{r^2\over a^2}\right)
\end{equation}
and
\begin{equation}
\label{equ:rho_2d_m}
T_{0}(r)= {e^{-{r^2\over2a^2}}\left[-2 a^4 e^{-{r^2\over2a^2}}+ \left(4a^4+2a^2 r^2+r^4\right)\right]\over 4\pi\left(a^3e^{-{r^2\over2a^2}}+a r^2\right)^2},
\end{equation}
and that of the higher-order multipoles
\begin{equation}
\label{equ:phi_2d_q}
\begin{split}
\Phi_{m}(r) = { {(1-e^{-{r^2\over2b^2}})}^{2m+1}\over r^m}
\end{split}
\end{equation}
and
\begin{equation}
\label{equ:rho_2d_q}
\begin{split}
T_{m}(r) = & {[\left(2m+1\right)e^{-{r^2\over2b^2}}{\left(1-e^{-{r^2\over2b^2}}\right)}^{2m-1}\over b^2 r^{m}]}\times\\
& \left[{2m r^2 e^{-{r^2\over2b^2}}\over b^2} - \left(1-e^{-{r^2\over2b^2}}\right)\left(2m-2+{r^2\over2b^2}\right)\right].
\end{split}
\end{equation}
In this work, we choose $a=50$ and $b=75$ for the 2D template profile. The Gaussian weighting in density in 2D and 3D is to render the density to approach zero rapidly. The unit of $a$ and $b$ is in pixel.
\label{lastpage}

\end{document}